# PROPERTIES OF THE OVERTONE OF THE ISOSCALAR GIANT MONOPOLE RESONANCE

M.L. Gorelik and M.H. Urin

*Moscow Engineering Physics Institute (State University), 115409 Moscow, Russia*

## Abstract

A semi-microscopic approach, based on the continuum-random-phase Approximation (CRPA) method, is applied to describe the main properties (strength function, transition density, direct-nucleon-decay branching ratios) of the overtone of the isoscalar giant monopole resonance. Results of the calculation are presented for $^{208}$Pb in comparison with those obtained for the isoscalar monopole and dipole giant resonances within the same approach.

Experimental and theoretical studies of highly-excited giant resonances (GR) have been undertaken in recent years to understand better how the different characteristics associated with GR formation (concentration of the particle-hole strength, coupling to the continuum, spreading effect) are affected with increasing GR energy. Most of the known highly excited GRs are the next vibration modes (the overtones) relative to the respective low-energy GRs (the main tones). The lowest energy overtone is the isoscalar giant dipole resonance (IS-GDR), studied intensively in recent experimental and theoretical works (see e.g. Refs. [1,2] and Refs. [3,4], respectively, and references therein). The ISGDR is the overtone of the $1^-$ zero-energy spurious state, associated with the center-of-mass motion. The isovector charge-exchange giant monopole and spin-monopole resonances are, respectively, the overtones of the isobaric analogue and Gamow-Teller giant resonances. The isovector monopole resonance, which plays an essential role in the isospin-mixing analysis (see e.g. Ref. [5]), has been identified in pion charge-exchange reactions [6]. The isovector spin-monopole resonance was recently observed in the ($^3$He,tp) and ($\vec{p},\vec{n}$) reactions (see Refs. [7] and [8], respectively).

One more candidate for studies of highly excited GRs is the overtone of the isoscalar giant monopole resonance, which can be related to the compression modes together with the ISGMR and ISGDR. From the microscopic point of view, this overtone (abbreviated below as ISGMR2) corresponds to $4\hbar\omega$ monopole particle-hole-type transitions, while the main tone corresponds to $2\hbar\omega$ transitions. The ISGMR2 was theoretically studied within a CRPA-based approach in Ref. [9] mainly for searching narrow ("trapped") resonances having a small relative strength. The ISGMR2 has been briefly described in Ref. [10], where the semi-classical approach and also the microscopic Hartree-Fock+RPA approach with the use of Skyrme interactions are employed.

In this paper, we describe the main properties of the ISGMR2 within the CRPA-based semi-microscopic approach. This approach has been previously used in Refs. [3,11] to describe the main properties of the ISGDR and ISGMR in several medium- and heavy-mass nuclei. (For brevity, we use below the notations of Refs. [3,11] and sometimes refer to equations from those references). The ingredients of the approach are the following: (i) the phenomenological (Woods-Saxon type) isoscalar part of the nuclear mean field and Landau-Migdal particle-hole (momentum-independent) interaction, (ii) the partial self-consistency conditions, used to restore within the RPA the translation invariance and isospin symmetry of the model Hamiltonian and also to calculate the mean Coulomb field and (iii) the phenomenological account for the spreading effect in terms of an energy-dependent smearing parameter. The choice of model parameters is given in Refs. [11,3] and briefly described below.

An important aspect of the theoretical studies of GR overtones within the RPA is the choice of an appropriate single-particle-probing operator (the external field). It is convenient to choose the operator with the condition that the main tone is not excited. In this case the overtone exhausts the most part of the respective particle-hole strength. The choice of the appropriate probing operator applied to description of the ISGDR, isovector monopole and spin-monopole giant resonances is discussed in Refs. [11], [12] and [13], respectively. To describe the ISGMR2 properties, we choose the radial part of the monopole probing operator $V^{(2)}_{L=0}$ in the form:

$$V^{(2)}_{L=0}(r) = r^4 - \eta_0 r^2; \quad \int \rho_{L=0}(r, \omega_{peak}) V^{(2)}_{L=0}(r) r^2 dr = 0. \tag{1}$$

Here, the parameter $\eta_0$ is determined by the integral condition, where $\rho_{L=0}(r, \omega_{peak})$ is the energy-dependent main-tone transition density taken at the peak energy of the ISGMR strength function. A similar probing operator has been used in Ref. [10].

Before turning to the results of the semi-microscopic description of the ISGMR2, we comment on the choice of model parameters used in calculations and also on the smearing procedure. The universal parameters of the isoscalar part of the nuclear mean field (including the spin-orbit term) and the isovector Landau-Migdal parameter $f' = 1.0$ have been used (see Eqs. (14)-(16) of Ref. [11]) to describe the nucleon separation energies for closed-shell subsystems in a wide mass interval (Table I of Ref. [11]). In this description, the isovector and Coulomb parts of the nuclear mean field are calculated selfconsistently via the neutron-excess and proton densities, respectively. The choice of the isoscalar Landau-Migdal parameters $f^{ex} = -2.897$ and $f^{in} = 0.0875$ allows us to bring the $1^-$ spurious-state energy close to zero and also to reproduce the experimental peak energy of the ISGMR in $^{208}$Pb [3]. The use of an energy-dependent smearing characterized by a saturation-like behavior:

$$I(\omega) = \alpha(\omega - \Delta)^2 / [1 + (\omega - \Delta)^2 / B^2], \tag{2}$$

with parameters $\alpha = 0.085$ MeV$^{-1}$, $\Delta = 3$ MeV, $B = 7$ MeV taken from Ref. [14], allows us to describe satisfactorily the experimental rms width of the isoscalar giant monopole and dipole resonances in a number of nuclei [3]. The smearing procedure comprises the replacement of the excitation energy $\omega$ by $\omega + \frac{i}{2} I(\omega)$ in CRPA equations to calculate the energy-averaged strength function and partial nucleon-escape amplitudes (the partial direct-nucleon-decay branching ratios). For this purpose, we calculate the $\omega$-dependent single-particle quantities (Green functions, continuum-state wave functions) using the single-particle potential having the imaginary part $\mp \frac{i}{2}I(\omega)f_{WS}(r, R^*, a)$. Here, the last factor is the Woods-Saxon function with $R^* > R$ ($R$ and $a$ are, respectively, the radius and diffuseness of the isoscalar part of the nuclear mean field). In our calculations we used $R^* = 1.3R$, since the energy-averaged effective probing operator (and, therefore, the strength function) is practically independent of $R^* > 1.3R$.

The results of the calculation for $^{208}$Pb are presented below. First, we found the parameter $\eta_0 = 75.6$ fm$^2$ from Eq. (1), where the transition density $\rho_{L=0}(r, \omega_{peak})$ is calculated by the CRPA-method via an equation similar to Eq. (1) of Ref. [3]. Then, we calculate the energy-averaged strength function $\bar{S}^{(2)}_{L=0}(\omega)$ and transition density $\bar{\rho}^{(2)}_{L=0}(r, \omega)$ with the use of the probing operator of Eq. (1). In Fig. 1, the reduced strength function $y_L(\omega) = \omega \bar{S}_L(\omega)/(EWSR)_L$ is shown for the ISGMR2 and compared with those calculated for the main tone and also for the ISGDR by the same method. As follows from Fig. 1, the ISGMR2 strength function is concentrated in three energy regions: the main peak and two "pygmy resonances" at lower energies.

The use of the probing operator of Eq. (1) is appropriate for describing ISGMR2. This statement is supported by an analysis of the relative strengths (fractions EWSR) $x_V(\omega) = \int_0^\omega \omega' S_V(\omega')d\omega'/(EWSR)_V$ calculated with the use of different monopole probing operators $V_{L=0}(r) : r^2, r^4, r^4 - R^2 r^2$ [9], $r^4 - \eta_0 r^2$ (Fig. 2). As it follows from the results, presented in Figs. 1 and 2, the main ISGMR2 component is at a relatively high excitation energy ($\omega_{peak} = 32$ MeV) and exhausts more than one half of the respective total strength. To compare the energy-dependent transition densities related to different energy regions, we have calculated the reduced energy-averaged densities $\mathcal{R}_V(r, \omega) = r^2 \bar{\rho}_V(r, \omega) \bar{S}^{-1/2}(\omega)$, normalized by the condition $\int \mathcal{R}_V(r, \omega)V(r)dr = 1$. The transition density $\mathcal{R}_{L=0}(r, \omega_{peak} = 14$ MeV) of the ISGMR (with the probing operator proportional to $r^2$) exhibits, naturally, one node inside the nucleus (Fig. 3). The radial dependence of this density is closely related to the ISGMR transition density $\rho_{L=0}(r)$ calculated in Ref. [11]. The reason is that the ISGMR strength function is well described within the CRPA by the single-level Breit-Wigner formula and, therefore, the radial and energy dependences of $\mathcal{R}_{L=0}(r,\omega)$ are factorized. As expected, the calculated radial dependence of the ISGMR2 transition density $\mathcal{R}_{L=0}(r,\omega_{peak} = 32$ MeV$)$ exhibits two nodes inside the nucleus (Fig. 4). This radial dependence is close to that of the ISGMR2 transition density found in the semi-classical approach of Ref. [10]. The radial dependence of $\mathcal{R}^{(2)}_{L=0}(r,\omega)$ is changed noticeably with decreasing the excitation energy (Fig. 4).

The respective energy-averaged nucleon-escape amplitudes (Eq. (2) of Ref. [3]) determine partial branching ratios for the GR direct nucleon decay. The branching ratios $b_\mu$ ($\mu^{-1}$ is the single-hole state populated after the decay), calculated for the ISGMR2 main peak, are given in Table 1 together with similar results for the ISGDR. The calculated $b_\mu$ values include the experimental spectroscopic factors $S_\mu$. The contribution of deep-hole states is also taken into account to evaluate the total branching ratio $b_{tot}$. In spite of about 10 MeV difference between the peak energies of the ISGDR and ISGMR2 (implying increased potential-barrier penetrability for nucleons escaping from ISGMR2), the evaluated total direct-neutron-decay branching ratios are found to be close in value for both resonances. Possibly, this fact is explained by the difference in the number of nodes in the transition-density radial dependences: one node for the ISGDR and two nodes for the ISGMR2. (Among all of the above-mentioned overtones only the ISGMR2 has such a transition density). We also note that the branching ratios, calculated for the ISGDR in the present work (Table 1), are noticeably smaller than those reported in Ref. [3]. This difference may be attributable to the use of the "non-smeared" $\omega$-dependent continuum-state wave functions in calculations of the energy-averaged partial nucleon-escape amplitudes in Ref. [3]. Although a similar procedure was used in Refs. [12] and [13] to calculate the direct-proton-decay branching ratios for the isovector monopole and spin-monopole giant resonances, respectively, the relative difference in the results is not large for these resonances due to their strong coupling to the continuum.

The relatively large value of the calculated total direct-proton-decay branching ratio for the ISMGR2 (Table 1) makes possible an experimental search for this resonance in the proton-decay reaction channels (e.g. in the $(\alpha, \alpha'p)$ reaction). A similar attempt to search for the isovector spin-monopole resonance in the $^{208}$Pb($^3$He,tp) reaction has proved successful [7].

In conclusion, we describe the main properties of the overtone of the isoscalar giant monopole resonance within the continuum-RPA-based semi-microscopic approach. A way to search experimentally for this overtone is also indicated.

The authors are grateful to U. Garg and M.N. Harakeh for interesting discussions and valuable remarks.

FIGURES

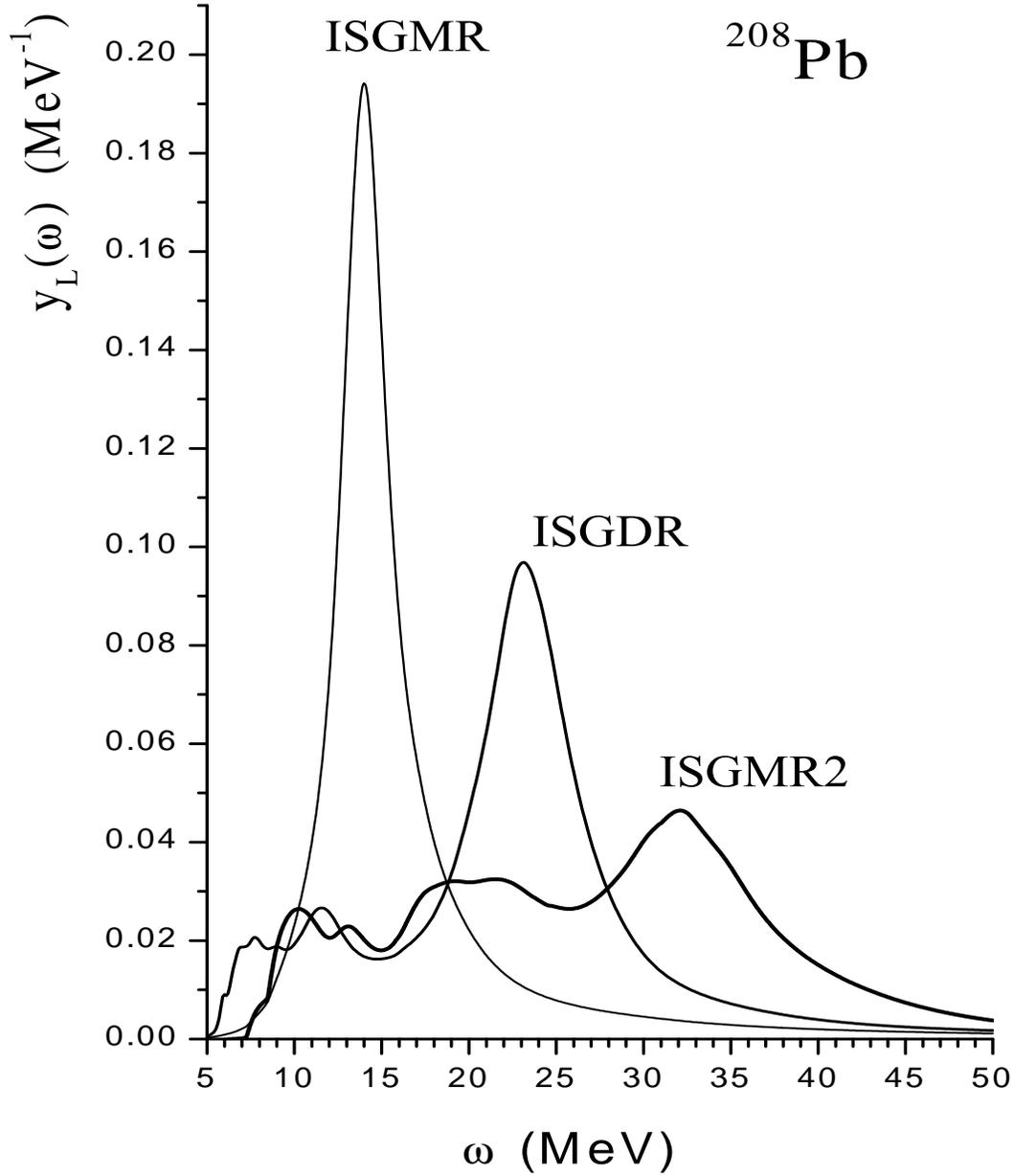

FIG. 1. The calculated relative strength functions of the isoscalar giant monopole and dipole resonances in $^{208}$Pb.

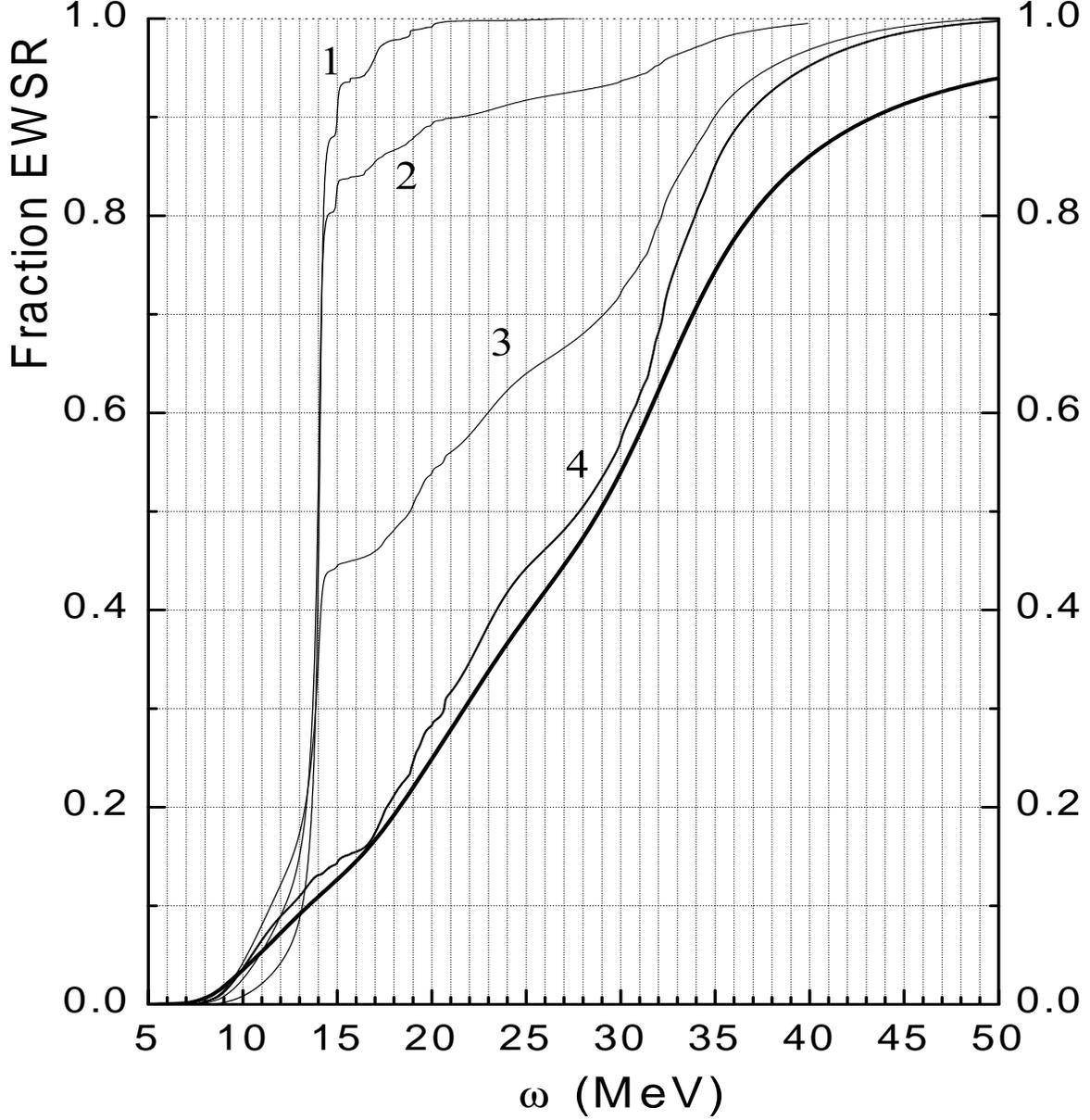

FIG. 2. The relative strength of monopole excitations in $^{208}$Pb for different excitation-energy intervals. The strength is calculated in CRPA (thin lines) with the use of some radial dependences for the monopole probing operator: $r^2(1), r^4(2), r^4-R^2r^2(3), r^4-\eta_0 r^2(4)$. The full line corresponds to calculations with taking the spreading effect into account.

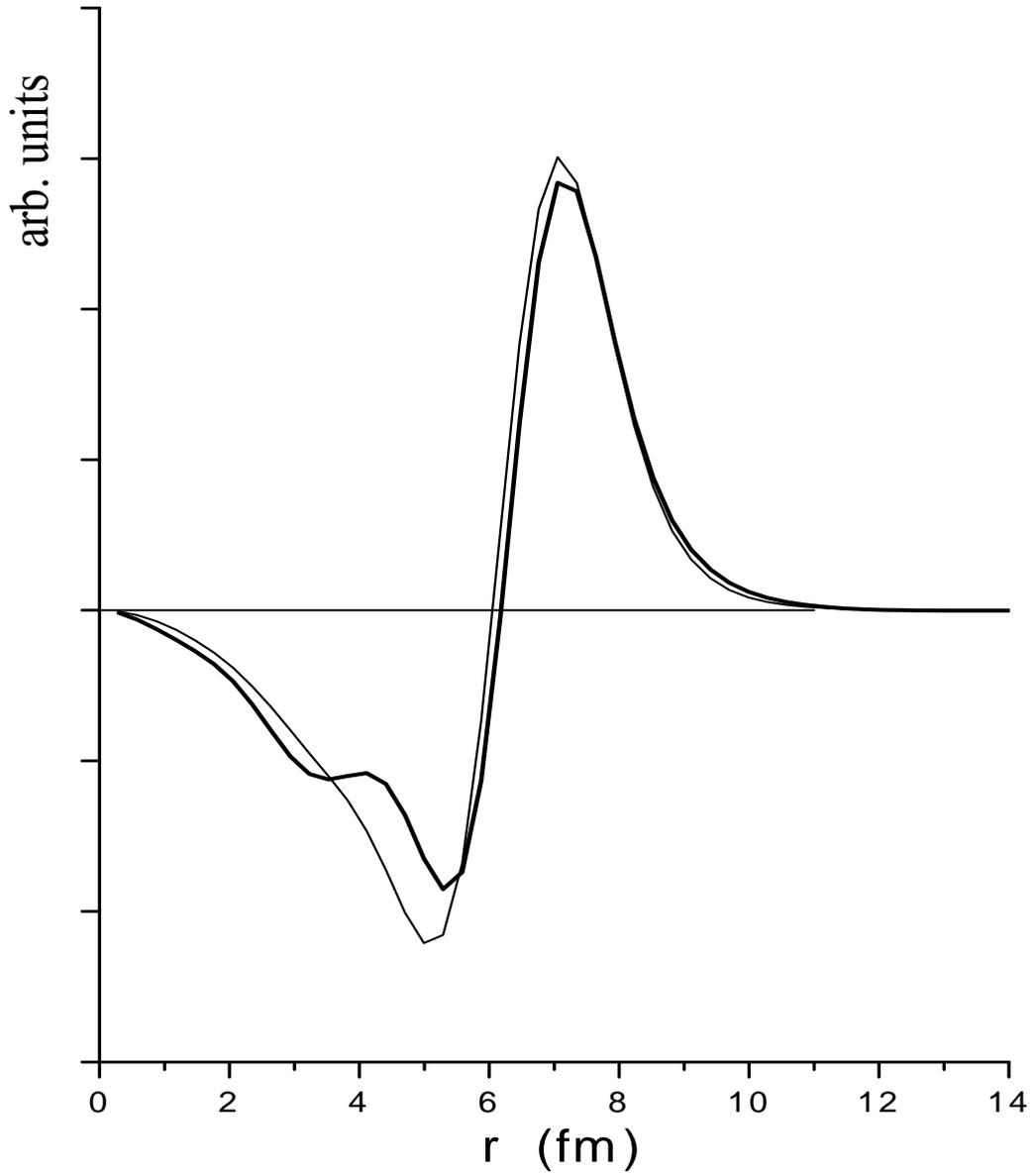

FIG. 3. The normalized energy-dependent transition density of the ISGMR in $^{208}$Pb calculated at the peak energy 14 MeV (the full line). The thin line corresponds to the collective ISGMR transition density calculated in the scaling model [17] and normalized in the same way.

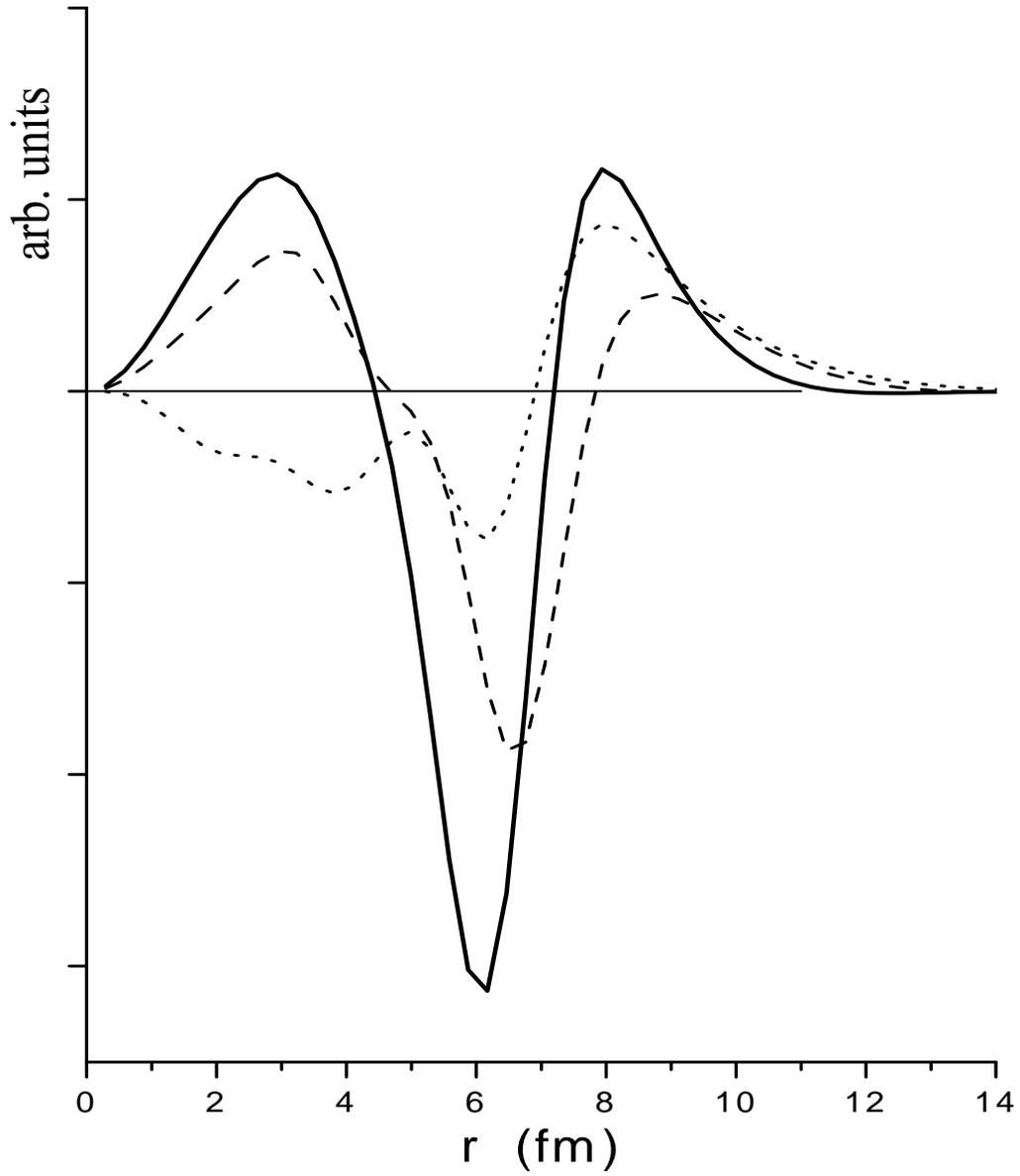

FIG. 4. The normalized energy-dependent transition density of the ISGMR2 in $^{208}$Pb calculated at several energies: 32 MeV (full line), 22 MeV (dashed), 12 MeV (dotted).

# REFERENCES

[1] H.L. Clark, Y.-W. Lui, and D.H. Youngblood, Phys. Rev. C **63**, 031301 (2001).

[2] M. Itoh et al., submitted to PRL.

[3] M.L. Gorelik and M.H. Urin, Phys. Rev. C **64**, 047301 (2001).

[4] S. Shlomo and A.I. Sanzhur, Phys. Rev. C **65**, 044310 (2002).

[5] N. Auerbach, Phys. Rep. **98**, 273 (1983).

[6] A. Erell et al., Phys. Rev. C **34**, 1822 (1986).

[7] R.G.T. Zegers et al., Phys. Rev. C **63**, 034613 (2001); R.G.T. Zegers et al., submitted to PRL.

[8] D.L. Prout et al., Phys. Rev. C **63**, 014603 (2001).

[9] S.E. Muraviev, I. Rotter, S. Shlomo, and M.H. Urin, Phys. Rev. C **59**, 2040 (1999).

[10] S. Shlomo, A.I. Sanzhur, and V.M. Kolomietz, Progress in Research, Cyclotron Institute, TAMU, April 1, 2000 - March 31, 2001, p.III-5.

[11] M.L. Gorelik, S. Shlomo, and M.H. Urin, Phys. Rev. C **62**, 044301 (2000).

[12] M.L. Gorelik and M.H. Urin, Phys. Rev. C **63**, 064312 (2001).

[13] V.A. Rodin and M.H. Urin, Nucl. Phys. **A687**, 276c (2001).

[14] V.A. Rodin and M.H. Urin, Phys. Lett. **B480**, 45 (2000).

[15] C.A. Whitten, N. Stein, G.E. Holland, and D.A. Bromley, Phys. Rev. **188**, 1941 (1969).

[16] I. Bobeldijk et al., Phys. Rev. Lett. **73**, 2684 (1994).

[17] S. Stringari, Phys. Lett. **108B**, 232 (1982).

TABLES

TABLE I. Calculated partial branching ratios for direct nucleon decay of the ISGDR and ISGMR2 in $^{208}$Pb. The results for decays with population of one-hole states from the last filled shells are shown for excitation-energy intervals 15-30 MeV (ISGDR) and 25-35 MeV (ISGMR2). Spectroscopic factors $S_\mu$ of these states are taken from Refs. [15] and [16] for neutrons and protons, respectively. $b_{tot}$ includes the contribution of deep-hole states. Spectroscopic factors $S_\mu = 0.6$ and $S_\mu = 0.5$ are taken for all deep-hole states in neutron and proton shells of $^{208}$Pb, respectively.

| neutron, $\mu^{-1}$ | $(1/2)^-$ | $(5/2)^-$ | $(3/2)^-$ | $(13/2)^+$ | $(7/2)^-$ | $(9/2)^-$ | | |
|---|---|---|---|---|---|---|---|---|
| $S_\mu$ | 1.0 | 0.91 | 0.98 | 1.0 | 0.7 | 0.61 | $\sum b_\mu^L$ | $b^{tot}$ |
| $b_\mu^{L=1}$ (%) | 0.7 | 2.2 | 1.8 | 3.2 | 2.9 | 1.0 | 11.8 | 15.5 |
| $b_\mu^{L=0}$ (%) | 0.2 | 0.6 | 0.6 | 1.0 | 1.2 | 0.5 | 4.1 | 13.3 |
| proton, $\mu^{-1}$ | $(1/2)^+$ | $(3/2)^+$ | $(11/2)^-$ | $(5/2)^+$ | $(7/2)^+$ | | | |
| $S_\mu$ | 0.55 | 0.57 | 0.58 | 0.54 | 0.26 | | | |
| $b_\mu^{L=1}$ (%) | 0.55 | 0.68 | 0.19 | 0.64 | 0.04 | | 2.1 | 2.1 |
| $b_\mu^{L=0}$ (%) | 1.3 | 2.2 | 3.0 | 3.4 | 0.7 | | 10.6 | 13.8 |